\documentclass[aps,prd,reprint,a4paper,showpacs,nofootinbib,superscriptaddress]{revtex4-2}

\usepackage{bbm}
\usepackage{amsmath}
\usepackage{multirow}
\usepackage{easybmat}
\usepackage{graphicx}
\usepackage{float}
\usepackage{amssymb}
\usepackage{mathrsfs}
\usepackage{subfigure}
\usepackage{amssymb}
\usepackage{hyperref}
\usepackage{epstopdf}

\usepackage[utf8]{inputenc}
\hypersetup{
    colorlinks=true,
    linkcolor=red,
    citecolor=blue,
}
\usepackage{color}
\usepackage[T1]{fontenc}
\usepackage{txfonts}

\begin{document}
 \newcommand{\bq}{\begin{equation}}
 \newcommand{\eq}{\end{equation}}
 \newcommand{\bqn}{\begin{eqnarray}}
 \newcommand{\eqn}{\end{eqnarray}}
 \newcommand{\nb}{\nonumber}
 \newcommand{\lb}{\label}

\title{Quasinormal modes in two-photon autocorrelation and the geometric-optics approximation}
\author{Wei-Liang Qian}
\email{Email address: wlqian@usp.br}
\affiliation{Center for Gravitation and Cosmology, College of Physical Science and Technology, Yangzhou University, Yangzhou 225009, China}
\affiliation{Escola de Engenharia de Lorena, Universidade de S\~ao Paulo, 12602-810, Lorena, SP, Brazil}
\affiliation{Institute for theoretical physics and cosmology, Zhejiang University of Technology, Hangzhou, 310032, China}

\author{Kai Lin}
\affiliation {Hubei Subsurface Multi-scale Imaging Key Laboratory, Institute of Geophysics and Geomatics, China University of Geosciences, Wuhan, 430074, China}

\author{Xiao-Mei Kuang}
\affiliation{Center for Gravitation and Cosmology, College of Physical Science and Technology, Yangzhou University, Yangzhou 225009, China}

\author{Bin Wang}
\affiliation{Center for Gravitation and Cosmology, College of Physical Science and Technology, Yangzhou University, Yangzhou 225009, China}
\affiliation{School of Aeronautics and Astronautics, Shanghai Jiao Tong University, Shanghai 200240, China}

\author{Rui-Hong Yue}
\affiliation{Center for Gravitation and Cosmology, College of Physical Science and Technology, Yangzhou University, Yangzhou 225009, China}

\begin{abstract}
In this work, we study the black hole light echoes in terms of the two-photon autocorrelation and explore their connection with the quasinormal modes. 
It is shown that the above time-domain phenomenon can be analyzed by utilizing the well-known frequency-domain relations between the quasinormal modes and characteristic parameters of null geodesics.
We found that the time-domain correlator, obtained by the inverse Fourier transform, naturally acquires the echo feature, which can be attributed to a collective effect of the asymptotic poles through a weighted summation of the squared modulus of the relevant Green's functions.
Specifically, the contour integral leads to a summation taking over both the overtone index and angular momentum. 
Moreover, the dominant contributions to the light echoes are from those in the eikonal limit, consistent with the existing findings using the geometric-optics arguments.
For the Schwarzschild black holes, we demonstrate the results numerically by considering a transient spherical light source.
Also, for the Kerr spacetimes, we point out a potential difference between the resulting light echoes using the geometric-optics approach and those obtained by the black hole perturbation theory. 
Possible astrophysical implications of the present study are addressed.

\end{abstract}

\date{Feb. 11th, 2022}

\maketitle

\section{Introduction}\label{section1}

The {\it shadow} of a black hole~\cite{agr-strong-lensing-shadow-review-01, agr-strong-lensing-shadow-review-02} is an extreme manifestation of the strong gravitational lensing effect in the vicinity of its horizon.
It can be evaluated by analyzing the properties of unstable fundamental photon orbits (FPO)~\cite{agr-strong-lensing-shadow-19}.
The latter is defined as periodic bound null geodesics, which fall back to the light ring and spherical orbit, respectively, in the cases of the Schwarzschild and Kerr metrics.
When perturbed, the trajectory of a light ray in question will slightly deviate from its originally bounded but unstable orbit, 
and subsequently, a photon will circulate the black hole a multitude of times before escaping to a distant observer and furnishing a pixel on the boundary of the black hole silhouette in his (her) local sky. 

While the black hole shadow determines the spatially asymptotic boundary of increasingly demagnified subrings, the notion of FPO also gives rise to another intriguing feature in the time domain, known as light {\it echo} or {\it glimmer}.
Indeed, for two adjacent light rays simultaneously emanated from the same source and consecutively captured by a static observer, they might orbit the hole a different number of times.
To be specific, the arrivals of the two light signals are separated by a time delay, which should be equal to a multiple of the orbital period.
Therefore, when a transient optically-thin radiation source intersects a closed orbit, one is led to the conclusion that different light travel times cause the image of the source to echo~\cite{agr-strong-lensing-correlator-03}.
The above heuristic arguments have been reinforced with more realistic simulations.
Indeed, in numerical calculations, echoes of the intensity in the image of a Keplerian hotspot were spotted, where the time delays between subsequent echoes were found to be essentially identical to the period of the closed orbits.
Moreover, consistent results have also been obtained for the autocorrelations of the measured signals~\cite{agr-strong-lensing-correlator-03, agr-strong-lensing-correlator-04}.

It is noted that the above approaches are essentially based on the properties of the null geodesics.
On the other hand, it is understood that the strong gravitational lensing of light rays around a black hole is simply the high-energy limit, namely, the geometric-optics counterpart, of the electromagnetic perturbations in the relevant black hole spacetime.
Physically, at such a limit, the typical scale of spatial variation of the wavefront, roughly the size of the black hole, is more significant than the wavelength~\cite{agr-strong-lensing-02, book-general-relativity-Wheeler}.
In this regard, the light echo signature derived using the properties of null geodesics is not expected to comprise any information that does not survive the geometric-optics approximation.
For instance, for a light ring orbit that lies in the equatorial plane, the longitudinal geodesic motion is perpendicular to the zenith degree of freedom.
Therefore, the resulting autocorrelations based on the null geodesics must be entirely irrelevant to the number of nodes of the waveform in the latitudinal direction.
In this context, from the perspective of the black hole perturbation theory, one might argue that it seems more appropriate to calculate the correlation function in terms of fields rather than light rays.

Indeed, the geometric-optics approximation is a venerable topic in the framework of general relativity~\cite{book-general-relativity-Wheeler, agr-analog-gravity-review-01}.
Among others, the specific relations between the quasinormal modes of field perturbations and the characteristic parameters of the corresponding null geodesics have been extensively studied in the literature~\cite{agr-qnm-review-02}. 
Following the initial works by Goebel~\cite{agr-qnm-geometric-optics-01} and Ferrari and Mashhoon~\cite{agr-qnm-Poschl-Teller-02}, the subject was elaborated by Cardoso {\it et al.}~\cite{agr-qnm-geometric-optics-02} for Schwarzschild metric, 
and then generalized to the Kerr spacetimes by Yang {\it et al.}~\cite{agr-qnm-geometric-optics-06}.
The above relations are derived in the eikonal limit using the WKB approximation~\cite{agr-qnm-WKB-01} by collecting the leading and sub-leading terms of $1/\ell$.
It was shown that the real part of the complex quasinormal frequencies of a specific field perturbation is related to the orbital and precession frequencies of the FPO.
The imaginary part, on the other hand, turns out to be related to the Lyapunov exponent of the orbit~\cite{book-dynamic-system-Massimo}.
In practice, they provide increasingly accurate estimation as one goes to larger angular momentum $\ell\gg 1$.

Therefore, it would be interesting to generalize the above studies and examine the time-domain light echoes using the black hole perturbation theory.
In particular, it would be meaningful to confirm the existence of light echos in the resulting correlator and, in particular, verify whether some particular features might emerge, which were absent from the geometric-optics approach.
A pertinent study in this regard has been recently initiated by Chesler {\it et al.}~\cite{agr-strong-lensing-correlator-12}.
In their work, the authors explore the coherent autocorrelation functions measured at a single telescope.
Specifically, a more straightforward setting is employed instead of studying electrodynamics sourced by fluctuating electric currents: the temporal two-particle correlation function is evaluated numerically for a massless scalar sourced by a stochastic field localized near a Schwarzschild black hole.
Light echoes were consequently observed in terms of the peaks attained at times equal to integer multiples of the photon orbit period.
These results are physically significant and consistent with the existing geometric-optic analysis.
Along this line of thought, in the present work, we extend the study further to consider a more realistic scenario: the electromagnetic field.
Besides, it is meaningful to understand analytically how the time-domain correlator, obtained via the inverse Fourier transform, acquires the light echo feature.
By analyzing the contour integral, it is observed that the relevant integral can be rewritten into a weighted summation of the squared modulus of the Green's functions, which is taken over both the overtone index and angular momentum.
Moreover, the primary characteristics of the time-domain correlator are essentially governed by the low-lying poles of the autocorrelation function along the real axis of the complex frequency.
The light echoes obtained numerically are explained from an analytic viewpoint by further employing the well-known relations between the quasinormal modes and null geodesics mentioned above.

The remainder of the present work is organized as follows.
In the next section, we explore the formulae concerning the two-particle autocorrelation function for the electromagnetic field and discuss their relation with black hole quasinormal modes.
In Sec.~\ref{section3}, to exam the light echoes in a more realistic configuration, we numerically investigate the time-domain two-photon autocorrelation by assuming a spherical radiation source.
The analytical analysis is presented in Sec.~\ref{section4} by scrutinizing the pole structure of the related Green's function.
The explanation is illustrated by a toy model, which demonstrates that it is not a single null geodesic but a collection of them, represented by a series of asymptotic poles, that eventually gives rise to the light echoes.
Furthermore, we elaborate on a few aspects, such as the physical implication of the eikonal limit and the generalization to an anisotropic source.
In Sec.~\ref{section5}, the case of Kerr black holes is briefly discussed, where we point out some subtlety in the resonance condition for light echoes in rotating black holes.
The last section is devoted to further discussions and concluding remarks.

\section{Two-photon correlation function in Schwarzschild black holes} \label{section2}

In this section, under the black hole perturbation theory framework, the main results on the two-photon autocorrelations are derived.
We first define the autocorrelation function for electromagnetic field in a similar fashion to that for the scalar field, recently introduced in~\cite{agr-strong-lensing-correlator-12}.
The two-particle correlation function can be formally solved using a generic Green's function.
By exploiting the spacetime symmetry, one shows that the latter can be further resolved by the decompositions using vector spherical harmonics.
However, the resulting radial equation for the relevant degree of freedom, inferred from the generic Green function, does not qualify as an equation for a univariable Green's function.
As a result, one adopts a second Green's function $\mathcal{G}$ and expresses the generic Green's function in terms of $\mathcal{G}$ and the vector spherical harmonics.
Subsequently, the resulting expression for the autocorrelation function is derived in terms of a summation w.r.t. different angular components.
Interestingly, we point out that $\mathcal{G}$ is nothing but the Green's function one usually defines for the black hole quasinormal modes.
For conciseness, we relegate most of the tedious mathematical details to the Appendices~\ref{appdxA} and~\ref{appdxB} and only cite the results in the main text.

The background spacetime of a $(3+1)$ Schwarzschild black hole takes the form 
\begin{equation}
ds^2 = -f dt^2 + \frac{1}{f} dr^2 + r^2 [d \theta^2 + \sin^2 \theta d\varphi^2], \ \ \ f = 1 - \frac{2 M}{r}
\end{equation}
where $M$ is the mass of the black hole.
The electromagnetic perturbations are governed by the Maxwell equation
\begin{equation}
{F^{\mu\nu}}_{;\nu} =  S^\mu , \label{Maxwell} 
\end{equation}
where the electromagnetic tensor $F_{\mu\nu} = A_{\nu,\mu}-A_{\mu,\nu}$ and the source is governed by the four-current $S^\mu=4\pi J^\mu$.
The equation of motion for the four-potential $A_\mu$ formally reads 
\begin{equation}
{\mathscr{O}}^{\mu\nu}\left(A_\mu(x)\right) = S^\nu(x) , \label{formal_EoM}
\end{equation}
where $x\equiv x^\nu=(t,r,\theta,\varphi)$, and the operator ${\mathscr{O}}^{\mu\nu}$ is linear in the four-vector space.

The field equation Eq.~\eqref{formal_EoM} is solved by 
\begin{equation}
A_\mu(y) = \int \sqrt{-g}G_{\mu\nu}(y, x)S^\nu(x) dx , \label{formal_Gsol}
\end{equation}
where $dx\equiv dtdrd\theta d\varphi$ and the retarded Green's function $G_{\mu\nu}(x, y)$ satisfies
\begin{equation}
{\mathscr{O}_x}^{\mu\rho}\left(G_{\mu\nu}(x, y)\right) = \frac{1}{\sqrt{-g}}{\delta^\rho}_\nu\delta(x-y) , \label{formal_Geq}
\end{equation}
where the subscript $x$ indicates that the operation is only on the coordinate variable $x$.

It is essential to note that the operator $\mathscr{O}$ is invariant w.r.t. the spatial rotation due to the spherical symmetry of the black hole metric. 
The symmetry can be exploited by decomposing both $A^\mu$ and $S^\mu$ into a suitable basis consisting of scalar and vector spherical harmonics~\cite{book-angular-momentum-Edmonds} as given by Eqs.~\eqref{Adecom} and~\eqref{Jdecom}.
As a result, the angular part of the equation of motion is entirely governed by the properties of the vector spherical harmonics, irrelevant to any specific detail of $\mathscr{O}$.

Consequently, the radial part of the master equations for the expansion coefficients carries the pertinent physics involved.
They are obtained by substituting the decomposition into Eq.~\eqref{Maxwell}.
Since different angular components are decoupled, the resulting spatiotemporal evolution of the perturbations can be viewed as a superposition of those of individual harmonics.
Moreover, as an Abelian gauge field, the electromagnetic field consists of only two free degrees of freedom, corresponding to the two polarization states.
By appropriate combinations of the above expansion coefficients, one arrives at two independent variables $\Psi_i$ ($i = 1, 2$) defined in Eq.~\eqref{Psi_master}.
Accordingly, from the four resultant radial equations one sorts out two independent master equations, which possess identical forms~\cite{agr-qnm-11}
\begin{eqnarray}
\frac{\partial^2}{\partial t^2}\Psi^{\ell,m}_i(t, r_*)+\left(-\frac{\partial^2}{\partial r_*^2}+V_\mathrm{RW}\right)\Psi^{\ell,m}_i(t, r_*)=S^{\ell,m}_i(t, r_*) ,
\label{master_eq_Psi}
\end{eqnarray}
where $V_\mathrm{RW}$ is the Regge-Wheeler potential
\begin{eqnarray}
V_\mathrm{RW}=f\left[\frac{\ell(\ell+1)}{r^2}+(1-{\bar{s}}^2)\frac{r_h}{r^3}\right] ,
\label{V_master}
\end{eqnarray}
where $r_*=r_*(r)\equiv r+r_h \ln\left(\frac{r}{r_h}-1\right)$ is the tortoise coordinate,
the horizon $r_h=2M$, and the spin ${\bar{s}}=1$ for photon so the second term in the bracket vanishes.
The explicit forms of the source terms $S_i$ given in Eq.~\eqref{S_master} are governed by the external current $J_\mu$.
Also, since the master equations for $\Psi_{1,2}$ are identical, their Green's functions, defined by replacing the r.h.s. by a point-like source, possess the same pole structure.
For the purpose of our present study, we will restrict outselves to one independent degree of freedom $\Psi_1$.
Also, similar to~\cite{agr-strong-lensing-correlator-12}, we assume a spherical incoherent source which satisfies
\begin{eqnarray}
\langle S(x)S^\dagger(y)\rangle= \frac{\chi(r)\zeta(t)}{\sqrt{-g}}\delta(x-y) ,\nonumber\\
\label{S_incorr}
\end{eqnarray}
where the functions $\chi$ and $\zeta$ determine the radial and temporal dependences of the source. 

Now we proceed to discuss the two-particle autocorrelation function of a field $\Psi(t, \mathbf{r})$, defined as
\begin{eqnarray}
C(t, \mathbf{r})\equiv \langle \Psi(t, \mathbf{r})\Psi^\dagger(0, \mathbf{r})\rangle .
\label{C_corr_Psi}
\end{eqnarray}
Here, $C(t, \mathbf{r})$ measures the correlation of the signal separated by time $t$ at a given spatial coordinate $\mathbf{r}$.
Besides the ensamble average for the source, an average for the relevant time duration $t_{\mathrm{win}}$ is also implied so that 
\begin{eqnarray}
\langle f(\Delta t)f(0)\rangle\equiv\frac{1}{t_{\mathrm{win}}}\int_{-t_{\mathrm{win}}/2}^{t_{\mathrm{win}}/2} dt f(t+\Delta t)f(t) . \nonumber
\end{eqnarray}

In practice, the Fourier transform of the correlator is usually more amenable
\begin{eqnarray}
\widetilde C(\omega, \mathbf{r})= \int dt C(t,\mathbf{r})e^{i\omega t}=\langle | \hat \Psi(\omega,\mathbf{r}|^2 \rangle ,
\label{Cf_corr}
\end{eqnarray}
where the amplitude $\hat \Psi(\omega,\mathbf{r})$ is given by the windowed Fourier transform of the field, namely, 
\begin{equation}
\hat \Psi(\omega,\mathbf{r}) \equiv \frac{1}{\sqrt{t_{\mathrm{win}}}}\int_{-t_{\mathrm{win}}/2}^{t_{\mathrm{win}}/2} dt \Psi(t,\mathbf{r}) e^{i \omega t} .\label{hat_Psi}
\end{equation}
We shall consider the limit $t_{\rm win} \to \infty$, in which case Eq.~\eqref{hat_Psi} approaches the Fourier transform, apart from an irrelevant factor.

For electromagnetic four-potential, the two-particle correlation function takes the general form
\begin{eqnarray}
C_{\mu\nu}(t, \mathbf{r})\equiv \langle A_\mu(t, \mathbf{r})A_\nu^\ast(0, \mathbf{r})\rangle .
\label{C_corr_A}
\end{eqnarray}
As discussed above, in the present study, we will focus on the degree of freedom related to $\Psi_1$.
Moreover, the calculation of Eq.~\eqref{C_corr_A} can be facilitated by employing the formal solution in Green's function, Eq.~\eqref{formal_Gsol}.
Also, by considering the incoherent source Eq.~\eqref{S_incorr}, one finds
\begin{eqnarray}
&C&(t, \mathbf{r}) \nonumber\\
&=& \int \sqrt{-g}dt' dr' d\theta' d\varphi' {G_1}(t, t', \mathbf{r}, \mathbf{r}') {G_1}^\dagger(0, t', \mathbf{r}, \mathbf{r}') \chi(r')\zeta(t') \nonumber \\
&=& \int \sqrt{-g}dt' dr' d\theta' d\varphi' {G_1}(t-t', \mathbf{r}, \mathbf{r}') {G_1}^\dagger(-t', \mathbf{r}, \mathbf{r}') \chi(r')\zeta(t') ,\nonumber\\
\label{C_corr_Psi1}
\end{eqnarray}
where $G_1$ is the Green function associated with $\Psi_1$.
As the metric is static, Green's function is translational invariant in time.
Subsequently, Green's function only depends on the time difference, and therefore one replaces $(t, t')$ by their difference.
Again, one resorts to its Fourier transform, which yields
\begin{eqnarray}
\widetilde C(\omega, \mathbf{r})=&& \int \sqrt{-g}d\omega' dr' d\theta' d\varphi' \nonumber\\
&&\times\ {\widetilde G_1}(\omega, \mathbf{r}, \mathbf{r}'){\widetilde G_1}^\dagger(\omega-\omega', \mathbf{r}, \mathbf{r}') \tilde{\zeta}(\omega')\chi(r') ,
\label{wC_corr_Psi1}
\end{eqnarray}
where $\tilde{\zeta}(\omega)$ is the Fourier transform of $\zeta(t)$.
The frequency-domain correlator can be further simplified by making use of the explicit form of the Green's function Eq.~\eqref{G1_ell_m_i} derived in the Appendix~\ref{appdxB} and orthonormal condition Eq.~\eqref{OrthoNormal_VSH}.
One finally obtains the desired result
\begin{eqnarray}
\widetilde C(\omega, \mathbf{r})=&& \sum_{\ell,m} \left|{{\mathscr V}_{(2)}^{\ell,m}}(\theta,\varphi)\right|^2 \int d\omega' {r'}^{-2}dr'  \nonumber\\
&&\times\ \mathcal{G}^{\ell,m}(\omega, r, r'){\mathcal{G}^{\ell,m}}^\dagger(\omega-\omega', r, r')\tilde{\zeta}(\omega')\chi(r')    \nonumber\\
=&&  \frac{1}{4\pi}\sum_{\ell} (2\ell+1) \int d\omega' {r'}^{-2}dr' \nonumber\\
&&\times\ \mathcal{G}^{\ell,m}(\omega, r, r'){\mathcal{G}^{\ell,m}}^\dagger(\omega-\omega', r, r')\tilde{\zeta}(\omega')\chi(r') ,
\label{wC_corr_Psi1_sum}
\end{eqnarray}
where $\ell$ and $m$ are the angular momentum and magnetic quantum number.
To derive it, the orthonormal relation Eq.~\eqref{OrthoNormal_VSH} and Uns\"old's theorem for vector spherical harmonics were utilized on the last line. 

It is important to note that $\mathcal{G}^{\ell,m}(\omega, r, r')$ satisfies Eq.~\eqref{master_radial_eq_G} with appropriate boundary conditions.
Therefore, by construction, it is precisely the Green's function of the radial master equation for the quasinormal modes.
Up to this point, we have established our main result of this section, which expresses the two-particle autocorrelation function in terms of Green's function of the black hole quasinormal modes.
In the frequency domain, the resultant correlator is expressed as a weighted sum of the squared modulus of the Green’s functions, where the summation is carried out w.r.t. different angular components. 
Moreover, the present case is different from the scenario where an initial pulse is planted to the system in order to assess the stability of the underlying black hole metric.
To be specific, a transient external source, rather than some initial condition, is assigned to the r.h.s. of the wave equation, which manifests itself in Eq.~\eqref{wC_corr_Psi1_sum} by the term $\tilde{\zeta}(\omega')\chi(r')$.

Now, one may also notice that the above derivation resides in the somewhat unrealistic assumption on the simple form of the incoherent source.
We will postpone discussions in this regard until Sec.~\ref{section4}, where we justify that the main features of the autocorrelation function, such as light echoes, should not depend on such simplifications.

In the next section, we proceed to evaluate the two-photon autocorrelation function Eq.~\eqref{C_corr_Psi} by considering a spherical source.
It is obtained numerically by the inverse Fourier transform wave function in the frequency domain.
As will be shown shortly, the resulting time-domain profile is characterized by light echoes.

\section{Light echoes from a transient spherical emission source} \label{section3}

\begin{figure}
\centering
\includegraphics[height=2.in,width=3.in]{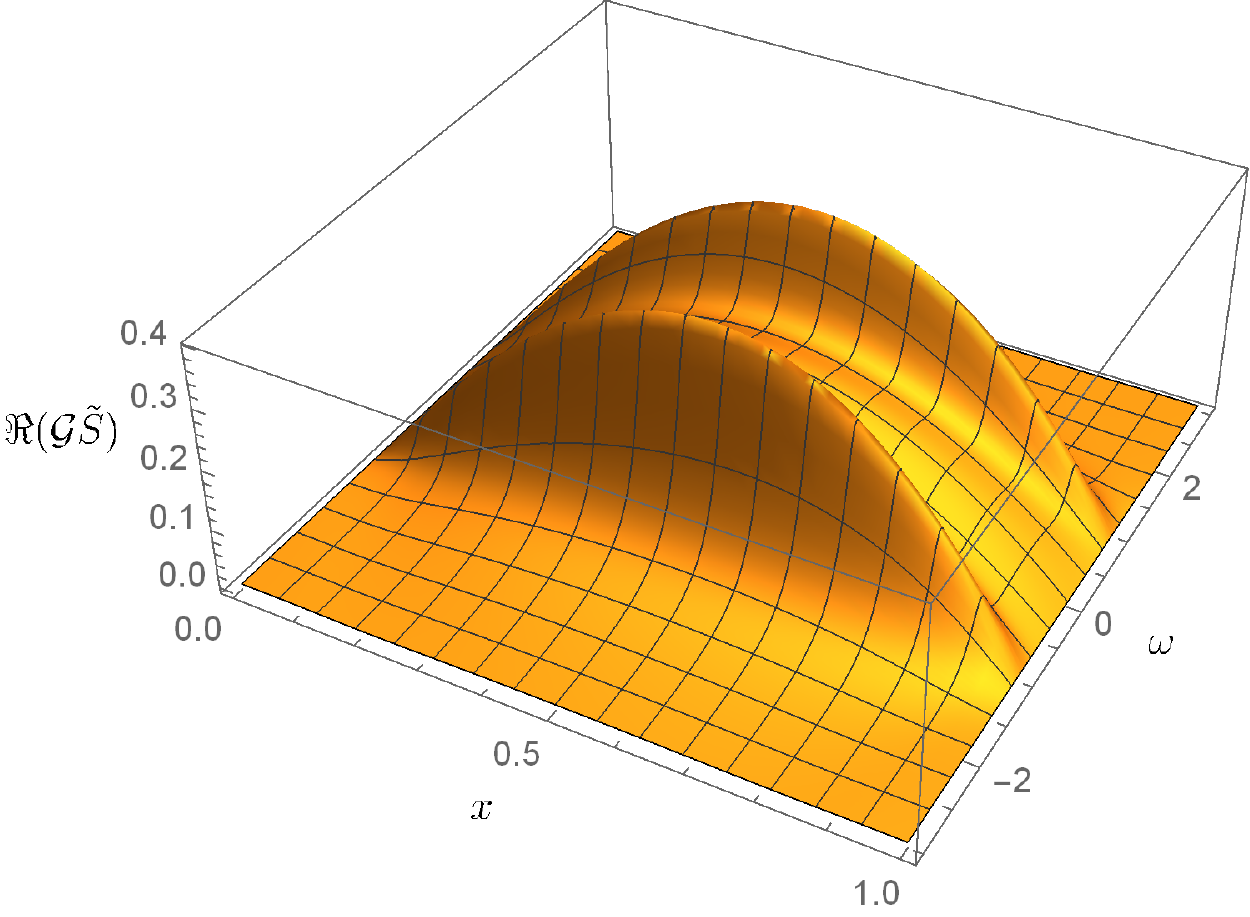}
\caption{(Color online) The frequency-domain Green's function integrated by the source $\int dr' \mathcal{G}(\omega, r, r')S(\omega, r')$, for a given angular momentum component $\ell=2$.
Here the radial coordinate is expressed in $\bar{x}=(r-r_h)/r$.}
\label{Gomega}
\end{figure}

In order to evaluate the autocorrelation function, we employed the scheme proposed in~\cite{agr-qnm-lq-02}.
We will relegate the detail of the numerical approach to Appendix~\ref{appdxC} and present the main idea and result in this section.

First, one obtains the correlator in the frequency domain for a given external source using the matrix method~\cite{agr-qnm-lq-matrix-01, agr-qnm-lq-matrix-02}.
The following source will be utilized
\begin{eqnarray}
S^{\ell,m}(\omega, r) = \frac{1}{\omega^2+1}e^{i\omega r}V_\mathrm{RW}(r) ,
\label{chi_choice}
\end{eqnarray}
where the factor $e^{i\omega r}V_\mathrm{RW}(r)$ is to ensure that the source satisfies appropriate boundary conditions.
The factor $\frac{1}{\omega^2+1}$ corresponds to a transient pulse in the time domain, which also effectively suppresses the numerical integral on the real axis of $\omega$.
The resultant frequency-domain wave function for a given angular momentum $\ell=2$ is shown in FIG.~\ref{Gomega}.
As pointed out in~\cite{agr-strong-lensing-correlator-12}, the Regge-Wheeler potential given by Eq.~\eqref{V_master} implies that there is a turning point at $\omega_c^2=f\frac{{\ell(\ell+1)}}{r^2}$.
The region $|\omega| <\omega_c$ is thus classically forbidden, and therefore, its contribution becomes exponentially small.
In FIG.~\ref{Gomega}, this corresponds to the valley alone $\omega\sim 0$.
As dictated by the Wronskian in the denominator~\cite{agr-qnm-review-02}, the Green's function approaches zero as $\mathcal{G}\sim \frac1\omega$ when $\omega\to\pm\infty$.
However, usually accompanied by strong oscillations, the convergence occurs slowly, which potentially poses a difficult task for numerical integration on the real axis.
Here, the apparent quick suppression shown in the plot is owing to the assumed factor $\frac{1}{\omega^2+1}$ from the frequency-dependent source.
Moreover, the stripes of maxima shown as the ``double shoulder'' in FIG.~\ref{Gomega} are related to the real part of the corresponding fundamental mode $\omega_{n=0,\ell=2}$.
This is expected since the nearest pole in the complex plane will largely affect the value of an analytic function on the real axis.
Apparently, such characteristics are attached to particular low overtone fundamental quasinormal mode.
Therefore, they are sensitively dependent on the spin of the perturbation field $\bar{s}$ as well as the specific metric in question.
As discussed above and will be further elaborated in the next section, the light echoes are {\it not} directly caused by these particular features.

\begin{figure}
\centering
\includegraphics[height=2.in,width=3.in]{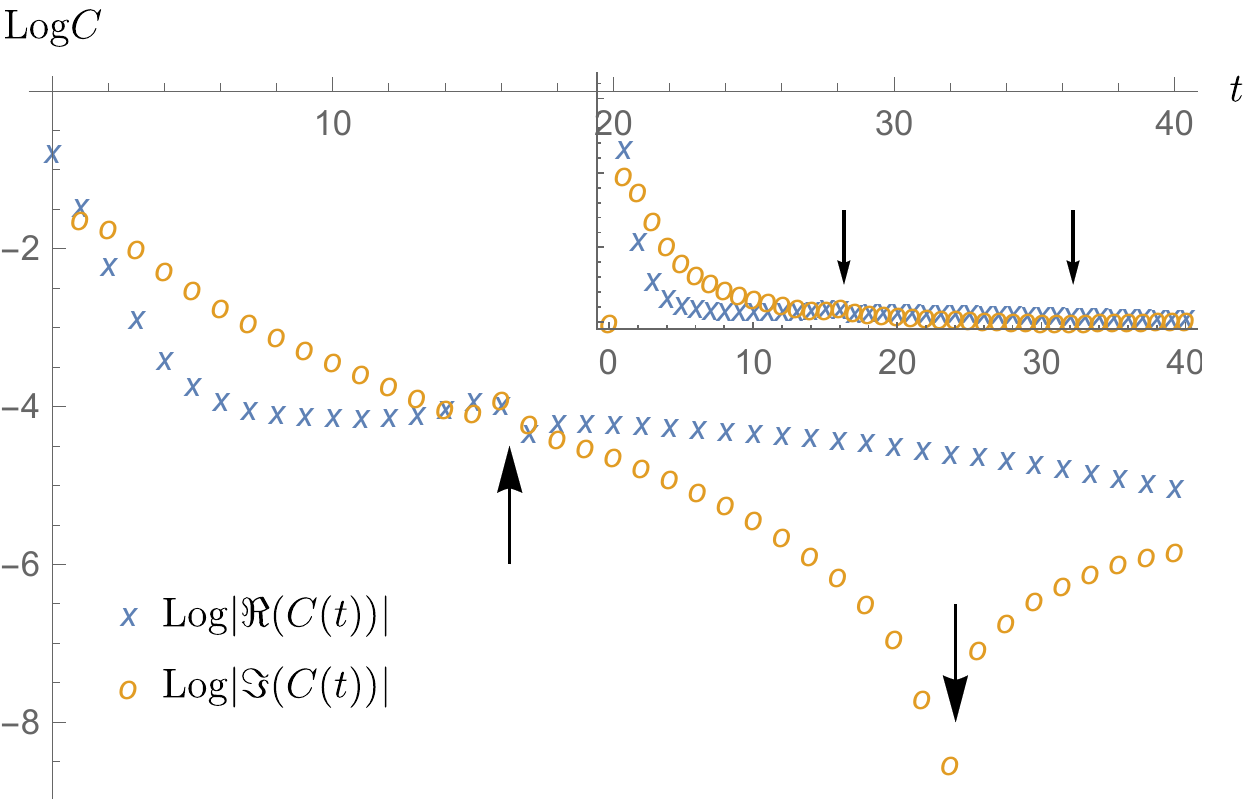}
\caption{(Color online) The calculated time-domain two-photon autocorrelation function.
The light echoes (indicated by black arrows) are observed at integer multiples of the light ring orbital period.
The inset shows the same results without using the logarithmic scale.}
\label{F_light_echoes}
\end{figure}

The resultant time-domain two-photon correlation function can be obtained through the momentum space waveform 
\begin{eqnarray}
\widetilde \Psi(\omega, \mathbf{r})= \sum_{\ell,m}{\mathscr V}_{(2)}^{\ell, m}(\theta, \varphi) \int dr' \mathcal{G}^{\ell,m}(\omega, r, r')S^{\ell,m}(\omega, r')  ,
\label{wC_Psi1}
\end{eqnarray}
which includes an summation over different angular components.
Subsequently, the correlator is obtained by numerically taking the inverse Fourier transform using Eq.~\eqref{Cf_corr}.
In our calculations, we assume that the observer sits at the north pole $\theta=\varphi=0$ at an arbitrary radial coordiante $r = 7.2$, and the term ${\mathscr V}_{(2)}^{\ell, m}(\theta, \varphi)$ is thus evaluated numerically.
The summation is evaluated for the angular components up to $\ell \le 20$.
In FIG.~\ref{F_light_echoes} we show the resultant two-particle autocorrelations.
It is observed that both functions peak at integer multiple of the light ring orbital period $T = \frac{2\pi r_\mathrm{LR}}{\sqrt{f}}=6\pi\sqrt{3}M= 3\pi\sqrt{3}$, where we have assumed $r_h=1$ in our calculations.
Physically, these peaks are identified to be the light echoes discussed in~\cite{agr-strong-lensing-correlator-12} for scalar perturbations and the black hole glimmer investigated in~\cite{agr-strong-lensing-correlator-03}.

In~\cite{agr-strong-lensing-correlator-12}, the light echoes in the two-particle autocorrelation function were also observed.
The calculations were carried out for the scalar perturbations, and the authors have adopted a different approximate approach.
In the following section, we elaborate on an explanation of the emergence of echoes by analyzing the properties of the Green's functions $\widetilde C(\omega, \mathbf{r})$ and $\mathcal{G}^{\ell, m}(\omega, r, r')$. 

\section{Collective behavior of asymptotic quasinormal modes at the eikonal limit} \label{section4}

\begin{figure}
\centering
\includegraphics[height=2.in,width=3.in]{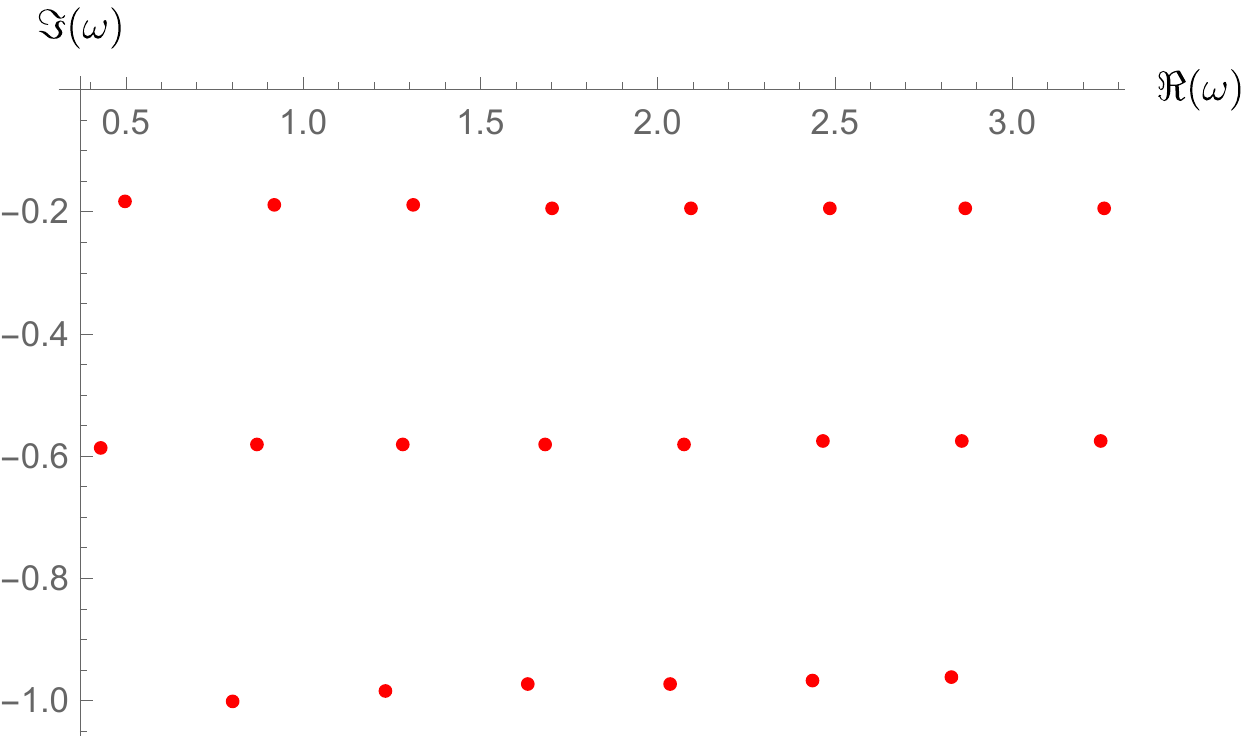}
\caption{(Color online) The quasinormal mode spectrum of electromagnetic perturbations for different overtone numbers $n$ and angular momenta $\ell$.
For the electromagnetic perturbations in the Schwarzschild black hole, an extensive study was performed in~\cite{agr-qnm-51}.}
\label{QNM_eikonal}
\end{figure}

In this section, we explore the origin of the light echoes in terms of analyzing the pole structures of the relevant Green's functions.
We argue that the resultant light echoes in the time domain can be attributed to a {\it collective effect} of the low-lying asymptotic quasinormal modes at the eikonal limit.
This is shown by primarily employing the well-known results in the literature about the relations between the quasinormal modes and null geodesics.

Formally, the time-domain autocorrelation function can be obtained by the inverse Fourier transform of Eq.~\eqref{Cf_corr}.
Using the specific form of Eq.~\eqref{wC_corr_Psi1_sum}, one has
\begin{eqnarray}
C(t, \mathbf{r})=&&\int d\omega e^{-i\omega t}\widetilde C(\omega, \mathbf{r})\nonumber\\
=&& \frac{i}{2}\sum_{n, \ell} e^{-i\omega_{n, \ell} t} (2\ell+1)  \mathcal{A}(\omega_{n, \ell}, t, r) \nonumber\\
&&\times\ \mathrm{Res}\left[\mathcal{G}^{\ell,m}(\omega, r, r'), \omega\right] ,
\label{wC_corr_time_sum}
\end{eqnarray}
where $n$ and $\ell$ are the overtone index and angular momentum, and $\omega_{n, \ell}$ corresponds to the spinor quasinormal frequency of the black hole metric.
The weight,
\begin{eqnarray}
\mathcal{A}(\omega_{n, \ell}, t, r)= \int d\omega' {r'}^{-2}dr' {\mathcal{G}^{\ell,m}}^\dagger(\omega_{n, \ell}-\omega', r, r')\tilde{\zeta}(\omega')\chi(r') ,\nonumber\\
\label{wC_corr_Aq}
\end{eqnarray}
results from the convolution of the initial condition and the conjugate of the Green's function ${\mathcal{G}^{\ell,m}}^\dagger$.
The relevant residues are those from $\mathcal{G}^{\ell,m}$.
This is because, according to Jordan's lemma, one should close the contour by an infinite semi-circle in the lower half-plane, which gives rise to a summation of the enclosed residues.
For a stable spacetime configuration, the poles of ${\mathcal{G}^{\ell,m}}$ lie below the real axis.
On the other hand, the complex conjugate mirrors all the poles of ${\mathcal{G}^{\ell,m}}$ w.r.t. to the real axis and, consequently, shifted along the real axis.
As a result, the latter will no longer be counted when evaluating the residues in Eq.~\eqref{wC_corr_Aq}, even though these reflected poles still affect the value of convolution in $\mathcal{A}$.
In other words, the relevant pole structure of $\widetilde C(\omega, \mathbf{r})$ and $\mathcal{G}^{\ell,m}(\omega, r, r')$ are identical.
As a result, the resultant correlator can be written as a summation, mainly proportional to the squared modulus of Green's function.
Moreover, the summation is taken over all possible overtone indices and angular momenta, $n$ and $\ell$ that, in principle, involves the entire quasinormal mode spectrum.

In fact, one can show that individual modes are manifestly present in the resultant time profile.
A straightforward way to see this is to numerically perform the integration on the first line of Eq.~\eqref{wC_corr_time_sum} and then try to extract the discrete frequencies using the Prony method~\cite{agr-qnm-55} from the time domain profile.
One may then compare the obtained complex frequencies against the values of the black hole quasinormal modes, as carried out in~\cite{agr-qnm-lq-02}.
By considering the term $\ell=2$ shown in FIG.~\ref{Gomega}, we numerically obtain the time-domain profile and then extract the most dominant modes.
The first two are numerically found to be $0.9169 -  0.193653 i$ and $0.889359 -  0.439647 i$, which are consistent with the values for the quasinormal modes of the lowest overtones, namely, $\omega_{n=0,\ell=2}=0.915191 - 0.190009 i$ and $\omega_{n=1,\ell=2}=0.873084 - 0.581421 i$.

It is appearant that the most dominant contributions in Eq.~\eqref{wC_corr_time_sum} are from the lowest-lying states with $n=0$, as the high-overtone ones are significant suppressed due to $e^{-i\omega_{n, \ell} t}$.
Moreover, the asymptotic distribution of quasinormal modes at the eikonal limit is featured by a uniform distribution~\cite{agr-qnm-Poschl-Teller-02, agr-qnm-geometric-optics-02, agr-qnm-geometric-optics-06}.
To be specific, for Schwarzschild black holes, one has
\begin{eqnarray}
\omega^{\mathrm{Sch}}_{n,\ell}\equiv\omega^{\mathrm{Sch}}_R+i\omega^{\mathrm{Sch}}_I 
=\left(\ell+\frac12\right) \Omega_{\mathrm{LR}}+i\left(n+\frac12\right)\gamma_{\mathrm{L}} , \label{QNM_LR_Schwarzschild}
\end{eqnarray}
where $\Omega_{\mathrm{LR}}$ and $\gamma_{\mathrm{L}}$ are the angular velocity and Lyapunov exponent of the light ring obit. 
In particular, the multipliers $\left(\ell+\frac12\right)$ and $\left(n+\frac12\right)$ in Eq.~\eqref{QNM_LR_Schwarzschild} indicate the poles are uniformly distributed parallel to the real axis, at large angular momentum number but for a given overtone number, as illustrated in FIG.~\ref{QNM_eikonal}.

To see intuitively how such uniform distribution of the quasinormal modes in the frequency domain gives rise to the time-domain light echoes, one devises the following toy model.
Let us construct a mathematically simple form of the Green function, which contains a series of poles governed by Eq.~\eqref{QNM_LR_Schwarzschild} given by
\begin{eqnarray}
\widetilde G(\omega)\sim \widetilde{F}(\omega)\widetilde{H}(\omega) ,
\label{G_factor}
\end{eqnarray}
where
\begin{eqnarray}
\widetilde F(\omega)\equiv \frac{1}{1 - e^{-i \omega T} B} ,
\label{F_factor}
\end{eqnarray}
and $T \in \mathbb{R}$ and $B \in \mathbb{C}$ are constants.
For our convenience, one further denotes $\Omega_{\mathrm{LR}} \equiv 2\pi/T$.

It is obvious that if $\omega_0  + \frac12 \Omega_{\mathrm{LR}}$ corresponds to one of its poles, then $\omega_0  + \left(j+\frac12\right) \Omega_{\mathrm{LR}}$ (for $j \in \mathbb{Z}$) must be another.
In other words, compared with the first term on the r.h.s. of Eq.~\eqref{QNM_LR_Schwarzschild}, these poles uniformly line up parallel to the real axis with an interval $\Omega_{\mathrm{LR}}$, thus the factor $F(\omega)$ mimics the asymptotic behavior at the eikonal limit.
The inverse Fourier transform of Eq.~\eqref{F_factor} gives
\begin{equation}
F(t) = \sqrt{2\pi}\left[\delta(t) + B \delta(t+T) + B^2 \delta(t+2T) + B^3 \delta(t+3T) + \cdots\right]  . \label{deltaSumFormal}
\end{equation}
As a result, the time-domain Green's function takes the form
\begin{equation}
\begin{split}
&{G}(t)\sim \int d\tau F(\tau)H(t-\tau) \\
&= \sqrt{2\pi}\left[ H(t) + B H(t+T) + B^2 H(t+2T)  + \cdots \right] , \label{deltaConvFormal}
\end{split}
\end{equation}
where $H(t)$ is the inverse Fourier transform of $\widetilde{H}(\omega)$.
Eq.~\eqref{deltaConvFormal} indicates that the resultant time-domain wave form is characterized by modulated pulses separated by the interval $T = 2\pi/\Omega_{\mathrm{LR}}$.
By Eq.~\eqref{QNM_LR_Schwarzschild}, it is nothing but the period of the light ring orbit, as predicted by the null geodesic analysis.

We note that one may readily use the residue theorem to recover the above result, which indicates the emergence of light echoes is independent of the specific function forms assumed by the toy model.
Besides, the essential part of the preceding derivation is not the absolute locations but the spacing between the poles.
This is in accordance with the fact that, for a static metric, the correlator Eq.~\eqref{wC_corr_time_sum} is translational invariance in time.
For instance, one may rewrite the frequency integral on the r.h.s. of Eq.~\eqref{wC_corr_Psi1_sum} as
\begin{eqnarray}
\int d\omega' \mathcal{G}^{\ell,m}(\omega-\omega', r, r'){\mathcal{G}^{\ell,m}}^\dagger(\omega, r, r')\tilde{\zeta}(\omega') .
\label{wC_corr_Psi1_sum_TRANS}
\end{eqnarray}
Now, the convolution $\int d\omega' {\mathcal{G}^{\ell,m}}(\omega-\omega', r, r')\tilde{\zeta}(\omega')$ shifts the poles of ${\mathcal{G}^{\ell,m}}$, with a weight determined by the external source $\tilde\zeta$, along the real axis.
However, the rest of the derivation that leads to the light echoes remains unchanged.

It is a good place to pause for a comment on the physical interpretation of the geometric-optics approximation.
In the framework of the present study, one may intuitively understand why the eikonal limit is in tune with a high-frequency radiation source.
To be specific, for a given high-frequency source, the momentum integration of Eq.~\eqref{wC_corr_Psi1_sum_TRANS} will receive more contributions from the region where the distribution $\tilde\zeta(\omega)$ is centered.
According to the above discussions, this is the same region where the quasinormal modes are equidistantly distributed, which eventually leads to more pronounced light echoes.
The arguments given above are not straightforwardly equivalent to the discussions from the null geodesic motivated analysis~\cite{agr-strong-lensing-correlator-03, agr-strong-lensing-correlator-04}.
For the latter, it is feasible that the light echoes are triggered by the perturbations of a {\it single} geodesic located on the light ring.
Indeed, the infinitesimal perturbation to a single geodesic on the light rings suffices to give rise to echoes. 
These geodesics are essentially not distinguishable because they carry very similar conserved numbers (e.g., energy, angular momentum in the $z$ direction, and Carter constant).

The source considered in Eq.~\eqref{S_incorr} is spherical, and a realistic radiation source is likely to be localized and therefore possesses other angular components.
By taking the latter into consideration, the angular part of Eq.~\eqref{wC_corr_Psi1_sum} will become more complicated.
To be specific, one has
\begin{eqnarray}
&&\sum_{\ell,m}{{\mathscr V}_{(2)}^{\ell, m}}^\dagger(\theta, \varphi){\mathscr V}_{(2)}^{\ell, m}(\theta, \varphi)  \nonumber\\
&\to& \sum_{\substack{\ell_1, \ell_2, \ell_3 \\ m_1, m_2, m_3}}{{\mathscr V}_{(2)}^{\ell_1, m_1}}^\dagger(\theta, \varphi){\mathscr V}_{(2)}^{\ell_2, m_2}(\theta, \varphi)S^{\ell_3, m_3}(\omega', r') \nonumber\\
&&\times\ \int d\Omega'\ {{\mathscr V}_{(2)}^{\ell_2, m_2}}^\dagger(\theta', \varphi'){\mathscr V}_{(2)}^{\ell_1, m_1}(\theta', \varphi'){\mathscr V}_{(2)}^{\ell_3, m_3}(\theta', \varphi') . \nonumber
\label{angularPart}
\end{eqnarray}
Subsequently, Eq.~\eqref{wC_corr_Psi1_sum} should be modified to read
\begin{eqnarray}
\widetilde C(\omega, \mathbf{r})=&& \int d\omega' {r'}^{-2}dr'  \mathcal{G}^{\ell,m}(\omega, r, r'){\mathcal{G}^{\ell,m}}^\dagger(\omega-\omega', r, r')\tilde{\zeta}(\omega')\chi(r')    \nonumber\\
&&\times\ \sum_{\substack{\ell_1, \ell_2, \ell_3 \\ m_1, m_2, m_3}}{{\mathscr V}_{(2)}^{\ell_1, m_1}}^\dagger(\theta, \varphi){\mathscr V}_{(2)}^{\ell_2, m_2}(\theta, \varphi)S^{\ell_3, m_3}(\omega', r') \nonumber\\
&&\times\ \int d\Omega'\ {{\mathscr V}_{(2)}^{\ell_2, m_2}}^\dagger(\theta', \varphi'){\mathscr V}_{(2)}^{\ell_1, m_1}(\theta', \varphi'){\mathscr V}_{(2)}^{\ell_3, m_3}(\theta', \varphi') .\nonumber\\
\label{wC_corr_Psi1_sum_mod}
\end{eqnarray}
In the place of the orthonormal relation, through which we once found an immediate simplification in deriving Eq.~\eqref{wC_corr_Psi1_sum}, one has to resort to the Clebsch-Gordan coefficients.
Also, Uns\"old's theorem becomes irrelevant.
The resulting factor for the angular part still turns out to be analytic but rather lengthy, as the summations in angular quantum numbers, and consequently, an explicit angular dependence, remain.
It is observed that the resulting angular dependence of the two-particle correlation function is expected from an anisotropic radiation source.
Now, instead of giving the explicit expression, we heuristically justify why such a complication will not affect our main results.
First, the coupling between different angular momenta will significantly simplify the summation in terms of the Clebsch-Gordan coefficients.
In other words, the angular factor will eventually furnish some combinatorial coefficients for the frequency-dependent sector of the correlator.
Therefore, the resulting correlator can still be viewed as a superposition of contributions from individual angular components.
Second, once the source possesses a spherical component ($\ell_3=m_3=0$), one always receives the corresponding contribution similar to that given by Eq.~\eqref{wC_corr_Psi1_sum}.

\section{Generalization to the Kerr black holes} \label{section5}

Owing to the sophisticated nature of its quasinormal modes~\cite{agr-qnm-geometric-optics-06} as well as null geodesics~\cite{agr-strong-lensing-07}, a full-fledged generalization to the Kerr black holes lies beyond the scope of the present study.
In this section, we make a preliminary attempt to address some of the key features of light echoes in rotating black holes.

From the geometric-optics arguments, a comprehensive analysis in this regard was recently given by Wong~\cite{agr-strong-lensing-correlator-03}.
In terms of null geodesics, the light echoes occur for specific FPOs whose trajectory intersects with itself.
For the Kerr black holes, the relevant null geodesics, namely, the spherical orbits, are usually {\it not} closed.
The resonance condition, which guarantees that the light ray eventually passes through the source again after a few revolutions, dictates that the ratio of the period of azimuthal motion to that of zenith one must be a rational number.
As discussed in~\cite{agr-strong-lensing-correlator-03}, for rotating black holes with arbitrary spin, it turns out that echoes always take place.
Specifically, there are infinitely many spherical orbits that meet the resonance condition, whose importance can be classified by the Stern-Brocot tree. 

From the viewpoint of the black hole perturbation theory, the physical interpretation of light echoes does not necessarily reside in the notion of closed null geodesics.
Specifically, based on the discussions in previous sections, the light echoes originate from the collective contributions from the asymptotic quasinormal modes.  
Moreover, as we argue below, light echoes do not always occur due to the quantized nature of quasinormal modes.

The black hole quasinormal modes for Kerr black holes at the eikonal limit were found~\cite{agr-qnm-geometric-optics-06} to possess the form
\begin{eqnarray}
\omega^{\mathrm{Kerr}}_{n, \ell, m}&\equiv&\omega^{\mathrm{Kerr}}_R+i\omega^{\mathrm{Kerr}}_I \nonumber\\
&=&\left[\left(\ell+\frac12\right)\Omega_{\theta}+ m \omega_{\mathrm{prec}}\right]+i\left(n+\frac12\right)\gamma_{\mathrm{L}} , \label{QNM_orb_Kerr}
\end{eqnarray}
where the angular velocity $\Omega_{\theta}=\frac{2\pi}{T_\theta}$ is defined in terms of $T_\theta$, the period of latitudinal oscillation in the zenith-angle $\theta$,
$\omega_{\mathrm{prec}}=\frac{\Delta\varphi_{\mathrm{prec}}}{T_\theta}$ gives the precession frequency in the azimuthal-angle $\varphi$,
and the Lyapunov exponent $\gamma_{\mathrm{L}}$ is defined by averaging over the period of a complete latitudinal oscillation.

By following the arguments given in Sec.~\ref{section4}, the relevant quasinormal modes are the fundamental ones ($n=0$).
Eq.~\eqref{QNM_orb_Kerr} indicates that the spectrum is governed by two indices, namely, $\ell$ and $m$. 
When compared against the analysis by the geometric-optics approach, the index $m$ can be associated with the role of the free parameter $r_0$, which denotes different spherical orbits.
It is straightforward to observe that the quasinormal spectrum will give rise to light echoes when the spectrum becomes degenerate, namely, the ratio between $\Omega_{\theta}$ and $\omega_{\mathrm{prec}}$ being a rational number.
Such a phenomenon of frequency degeneracy in the spectrum was first pointed out by Yang {\it et al.} in~\cite{agr-qnm-geometric-optics-06}.
Here we proceed further to argue that such degeneracy furnishes the resonance condition for the light echoes.
To be specific, when the above condition is satisfied for rotating black holes with a specific spin, light echoes will only occur since the quasinormal spectrum is then featured by a uniform distribution in the eikonal limit.
On the other hand, when the resonance condition fails, the distribution of the spectrum will become irregular, and therefore the arguments which lead to the light echoes are no longer valid.
We note that the latter usually is the case for a rotating black hole of arbitrary spin.
Here, the essential difference between the two rationalizations is that $r_0$ changes continuously while $m$ is quantized.  
In other words, such a distinctive feature is naturally attributed to the discrete nature of quasinormal frequencies.
Also, the different feature in light echoes is in accordance with what was pointed out in~\cite{agr-qnm-geometric-optics-06}, that a quasinormal mode can be mapped to a null geodesic, but not vice versa.

\section{Further discussions and concluding remarks} \label{section6}

The studies~\cite{agr-strong-lensing-correlator-03, agr-strong-lensing-correlator-04} based on geometric-optics indicated that black hole glimmer, or light echoes, associated with the periodic bound geodesics, can be used to infer important properties of the underlying compact object. 
On the other hand, in terms of the black hole perturbation theory, one can tackle the same problem via the two-particle autocorrelation of the electromagnetic field while adopting appropriate geometric-optics approximation.
As it is commonly understood, geometric optics is implied when the wavelength of the radiation is sufficiently small compared to the scale of the black hole.
Therefore, intuitively, the results on light echoes are expected to be deducible from the pole structure of Green's function defined in the context of the black hole perturbation theory.
The present study was motivated by the above considerations and aimed to clarify a few aspects.
While reinforcing the recent findings reported in~\cite{agr-strong-lensing-correlator-12}, we investigate the connection between the two approaches by focusing on the role of the geometric-optics approximation.
By analyzing the pole structure on the complex plane and employing the well-known relations between black hole quasinormal modes and null geodesics, we argue that the echo of the black hole image can be explained as a collective effect of low-lying quasinormal excitations in the eikonal limit.
In this context, the mechanism of the light echoes is reminiscent of how regular distribution of lattice structure leads to quantized wave vectors, known as the {\it Bloch wave}.
For Kerr black holes, light echoes occur when the resonance condition is satisfied.
Unlike the results from geometric-optics approaches, we pointed out that light echoes are most pronounced for rotating black holes of a particular spin.
It is understood that such a distinct feature is due to quasinormal frequencies' quantized nature.

As mentioned in the introduction, the relations between black hole quasinormal modes and null geodesics have been extensively explored in the literature.
In this work, we have pushed forward along this line of research and further extended such investigation to the time-domain two-particle autocorrelation functions.
The geometric-optics approximation plays an essential role in bridging two physically pertinent approaches: the black hole perturbations in terms of fields and the geometric-optics in terms of null geodesics.
The topic regarding the optics limit of field perturbations in terms of the Green's function was also investigated by Nambu and Noda~\cite{agr-strong-lensing-correlator-08}, where analysis was carried out for the massless scalar field in Schwarzschild spacetime.
In the study, Green's function is represented by the sum over the partial waves, which is, in turn, approximated by the residues of the Regge poles.
Besides the type of perturbations and the spacetime background, the main difference is that Ref.~\cite{agr-strong-lensing-correlator-08} primarily focused on the spatial dependence of the waveform at a given frequency.
We note that the mechanism of light echoes discussed in the present study is somewhat similar to a model recently proposed for gravitational wave echoes~\cite{agr-qnm-echoes-20}.
However, as the latter concerns mostly the high overtone modes of a given angular momentum~\cite{agr-qnm-lq-03}, we understand that the underlying physics of the two studies is rather different.
For the case of gravitational perturbations, the merger taken place for a more significant system does not necessarily emanate gravitational waves of higher frequency.
Therefore, the low-frequency signals, aimed by various ongoing space-borne detector projects, might be more relevant for gravitational radiations.
On the other hand, it is plausible for electromagnetic radiation to be emanated from a high-frequency source. 
As a result, the image and the glimmer of a black hole could be captured and reasonably described in terms of a geometric-optics approach.

It is also worth noting the particular case pointed out by Konoplya and Stuchl\'ik~\cite{agr-qnm-geometric-optics-14} for which the relation Eq.~\eqref{QNM_LR_Schwarzschild} is violated.
This seems to lead to the following dilemma that deserves further investigation. 
On the one hand, light echoes are naturally implied as long as bound unstable null geodesic exists.
On the other, while the distribution of the poles is still regular, it apparently indicates a resonance period ``incompatible'' with that deduced from the light ring orbit.

Recently, many exciting features of the black hole shadow have received much attention.
Intriguing speculations include cuspy~\cite{agr-strong-lensing-shadow-10, agr-strong-lensing-shadow-35}, fractured~\cite{agr-strong-lensing-shadow-35}, and open~\cite{agr-strong-lensing-shadow-36} shadows.
In particular, the black hole shadow has also been analyzed from the viewpoint of dynamical system~\cite{agr-strong-lensing-shadow-33}.
Specifically, the physically pertinent FPOs constitute a particular family of periodic bound orbits, which subsequently inherit the stability structure of the fixed point from which they emanate.
Moreover, the unstable manifold attached to them, in turn, furnishes the boundary of the black hole silhouette.
It is intriguing whether such features can be revisited from the perspective of the black hole perturbation theory.

Last but not least, we make a short comment on the possible astrophysical relevance of the black hole echoes.
Based on the {\it very long baseline interferometry} technique, the spatial correlations of the complex electromagnetic fields have been recorded by an array of synchronized millimeter telescopes.
The latter effectively acts as a single giant virtual telescope, namely, the Event Horizon Telescope~\cite{agr-strong-lensing-EHT-L01, agr-strong-lensing-EHT-L04, agr-strong-lensing-EHT-L05}, whose aperture is nearly the same as the diameter of Earth. 
As the culmination of decades-long efforts, the first released black hole image indicated that a novel avenue had been opened up for direct observation of the black holes in the electromagnetic channel.
As pointed out by some authors~\cite{agr-strong-lensing-correlator-03,agr-strong-lensing-correlator-04,agr-strong-lensing-correlator-12}, in the light of progressively better images of more black holes, the observation of black hole glimmer might become feasible, and therefore it is rather inviting for detailed predictions on the theoretical side.
In connection with the empirical observations, it would be of interest to further analyze the following aspects along the line of the present study. 
First, we have not investigated the effects of the relative location among the emission source, the black hole, and the observer.
For instance, for a unidirectional radiation source located on the equatorial plane, it was found~\cite{agr-strong-lensing-correlator-03, agr-strong-lensing-correlator-04} that the correlation function of the intensity fluctuations peaks at half of the light ring orbit period.
Also, if the observer sits at the north pole, most modes with nonvanishing magnetic quantum numbers will not be accessible, which, in turn, will significantly modify the observed correlation strength.
Second, the two-particle correlation function investigated in the present work is primarily related to the flux density, but not their correlations. 
A direct comparison with the latter quantity in the relevant studies would require evaluating the four-particle correlations.
Also, the subtle difference in the light echoes in Kerr black hole potentially leads to nontrivial implications on the experimental side.
We plan to explore these subjects further in future studies.

\section{Appendix} \label{appendix}
\renewcommand{\theequation}{A.\arabic{equation}}
\setcounter{equation}{0}

In Appendices~\ref{appdxA} and~\ref{appdxB}, we give a more detailed account of the decomposition of the electromagnetic four-potential in terms of the scalar and vector spherical harmonics, as well as derive the expressions for the Green's functions following essentially Refs.~\cite{agr-qnm-11, agr-strong-lensing-correlator-12}.
In Appendix~\ref{appdxC}, we present the numerical approach given in Sec.~\ref{section3}, where the scheme~\cite{agr-qnm-lq-02} is based on the matrix method~\cite{agr-qnm-lq-matrix-01, agr-qnm-lq-matrix-02}.

\subsection{Electromagnetic perturbations in Schwarzschild metric with external source}\label{appdxA}

The equation of motion of the electromagnetic perturbation can be formally given by Eq.~\eqref{formal_EoM}
\begin{equation}
{\mathscr{O}}^{\mu\nu}\left(A_\mu(x)\right) = S^\nu(x) ,\nonumber
\end{equation}
where the operator $\mathscr{O}^{\mu\nu}$ is invariant w.r.t. the spatial rotation, namely,
\begin{equation}
{\left(M^{-1}\right)^{\sigma}}_\mu{\mathscr{O}}^{\mu\nu}{M_{\nu}}^\rho = {\mathscr{O}}^{\sigma\rho} , 
\end{equation}
where ${M_\mu}^\nu$ is the matrix representation of the SO(3) group, which is a {\it little group} of the Lorentz group SO(1,3).
For the specific case at hand, the operation transforms the indices of the vector field according to the direct sum $D^0\oplus D^1$, while the coordinate arguments rotate reversely~\cite{book-qft-Weinberg-v1}.
By construction, $M$ is reducible.
The invariance Eq.~\eqref{formal_EoM} indicates that it would be beneficial to decompose both the perturbation and source into relevant bases in accordance with irreducible representations of the SO(3) group.
Besides the scalar functions, which are intuitively chosen to be the spherical harmonics $Y^{\ell,m}(\theta,\varphi)$, one may also introduce the following two vector spherical harmonics
\begin{eqnarray}
\left({\mathscr V}_{(1)}^{\ell,m}\right)_a&=& \frac{1}{\sqrt{\ell(\ell+1)}}\left({\mathscr S}^{\ell,m}\right)_{;a}=\frac{1}{\sqrt{\ell(\ell+1)}}\frac{\partial}{\partial x^a}Y^{\ell,m}(\theta,\varphi) , \nonumber \\ 
\left({\mathscr V}_{(2)}^{\ell,m}\right)_a&=& \frac{1}{\sqrt{\ell(\ell+1)}}{\epsilon_a}^b\left({\mathscr S}^{\ell,m}\right)_{;b}=\frac{1}{\sqrt{\ell(\ell+1)}}\epsilon_{ac}\gamma^{cb}\frac{\partial}{\partial x^b}Y^{\ell,m}(\theta,\varphi)  \nonumber,\\
\label{vharm}
\end{eqnarray}
where we have adopted the notation utilized in~\cite{agr-qnm-review-02}.
Here, $x^a=(\theta,\varphi)$ with the indices $a, b, c$ take $2$ or $3$, 
\begin{eqnarray}
\gamma_{ab}=\begin{pmatrix}1&0\\0&\sin^2\theta\end{pmatrix} 
\end{eqnarray}
is the projected metric and 
\begin{eqnarray}
\epsilon_{ab}=\sin\theta\begin{pmatrix}0&-1\\1&0\end{pmatrix} 
\end{eqnarray}
is the totally antisymmetric tensor on the two-sphere.
On the one hand, by construction, these quantities transform as vector fields as required.
Moreover, it is straightforward to observe that they furnish representations of the SO(3) group, which can be further shown to be irreducible~\cite{book-angular-momentum-Edmonds}.
On the other hand, an arbitrary four-vector $B_\mu$ accomodates two scalar components and one vectorial piece, namely,
\begin{equation} 
B_\mu(t,r,\theta,\varphi) = 
\left(
\begin{array}{c@{}c@{}c}
\begin{BMAT}{|c|}{|c|c|ccc|}
s \\
s \\
\vspace{1pt} \\
v  \\
\vspace{1pt}
\end{BMAT}  &&
\end{array}
\right) .
\end{equation}
Therefore, the most general form of decomposition of a four-vector $B_\mu$ consists of four terms.
Two coefficients are for the scalar ones associated with the time and radial components, namely, 
\begin{eqnarray}
\left({\mathscr S}_{(1)}^{\ell,m}\right)_\mu=\delta_{\mu 0}Y^{\ell,m}(\theta,\varphi) , \nonumber \\
\left({\mathscr S}_{(2)}^{\ell,m}\right)_\mu=\delta_{\mu 1}Y^{\ell,m}(\theta,\varphi) . \label{sharm}
\end{eqnarray}
Another two coefficients constitute a linear combination of the two vector spherical harmonics Eq.~\eqref{vharm} related to the angular part.
It is readily verified the four basis-harmonics defined above are orthonormal
\begin{eqnarray}
\int {{\mathscr S}_{(i)}^{\ell,m}}^\dagger(\theta,\varphi){{\mathscr S}_{(j)}^{\ell',m'}}(\theta,\varphi) d\Omega&=&\delta_{ij}\delta_{\ell,\ell'}\delta_{m,m'} , \nonumber \\
\int {{\mathscr V}_{(i)}^{\ell,m}}^\dagger(\theta,\varphi){{\mathscr V}_{(j)}^{\ell',m'}}(\theta,\varphi) d\Omega&=&\delta_{ij}\delta_{\ell,\ell'}\delta_{m,m'} ,\nonumber \\
\int {{\mathscr S}_{(i)}^{\ell,m}}^\dagger(\theta,\varphi){{\mathscr V}_{(j)}^{\ell',m'}}(\theta,\varphi) d\Omega&=& 0 .
\label{OrthoNormal_VSH}
\end{eqnarray}
It is important to note that the above four-component harmonic bases only represent the factorized angular part of a covariant four-vector, but they are not covariant themself.
For instance, ${{\mathscr V}_{(i)}^{\ell,m}}^\dagger$ indicates a row matrix of complex functions, as the conjugate transpose of ${{\mathscr V}_{(i)}^{\ell,m}}$.

Following the convention in~\cite{agr-qnm-11}, one denotes the coefficients for $A^\mu$ and $S^\mu$ by $(f, h, a, k)$ and $(\psi, \eta, \alpha, \chi)$, respectively. 
The resultant decompositions read
\begin{eqnarray}
A_\mu(t,r,\theta,\varphi)&=&\sum_{\ell,m} \left[f^{\ell,m}(t,r){\mathscr S}_{(1)}^{\ell,m}(\theta,\varphi)+h^{\ell,m}(t,r){\mathscr S}_{(2)}^{\ell,m}(\theta,\varphi)\right. \nonumber \\
&& \left.+k^{\ell,m}(t,r){\mathscr V}_{(1)}^{\ell,m}(\theta,\varphi)+a^{\ell,m}(t,r){\mathscr V}_{(2)}^{\ell,m}(\theta,\varphi)\right]\nonumber \\
&=&\sum_{\ell,m} 
\begin{pmatrix}
0\\
0\\
\frac{a^{\ell,m}(t,r)}{\sqrt{\ell(\ell+1)}}\frac{1}{\sin\theta}\frac{\partial Y^{\ell,m}(\theta,\varphi)}{\partial \varphi}\\
-\frac{a^{\ell,m}(t,r)}{\sqrt{\ell(\ell+1)}}\sin\theta\frac{\partial Y^{\ell,m}(\theta,\varphi)}{\partial \theta}
\end{pmatrix}
+\begin{pmatrix}
f^{\ell,m}(t,r) Y^{\ell,m}(\theta,\varphi)\\
h^{\ell,m}(t,r) Y^{\ell,m}(\theta,\varphi)\\
\frac{k^{\ell,m}(t,r)}{\sqrt{\ell(\ell+1)}}\frac{\partial Y^{\ell,m}(\theta,\varphi)}{\partial \theta}\\
\frac{k^{\ell,m}(t,r)}{\sqrt{\ell(\ell+1)}}\frac{\partial Y^{\ell,m}(\theta,\varphi)}{\partial \varphi}
\end{pmatrix}  \nonumber \\
\label{Adecom}
\end{eqnarray}
and
\begin{eqnarray}
J_\mu(t,r,\theta,\varphi)&=&\sum_{\ell,m} \left[\psi^{\ell,m}(t,r){\mathscr S}_{(1)}^{\ell,m}(\theta,\varphi)+\eta^{\ell,m}(t,r){\mathscr S}_{(2)}^{\ell,m}(\theta,\varphi)\right. \nonumber \\
&& \left.+\chi^{\ell,m}(t,r){\mathscr V}_{(1)}^{\ell,m}(\theta,\varphi)+\alpha^{\ell,m}(t,r){\mathscr V}_{(2)}^{\ell,m}(\theta,\varphi)\right] .\nonumber \\
\label{Jdecom}
\end{eqnarray}

By substituting Eqs.~\eqref{Adecom} and~\eqref{Jdecom} into Eq.~\eqref{Maxwell}, one obtains the equations of motion for the four expansion coefficients $(f, h, a, k)$.
As shown in~\cite{agr-qnm-11}, the current conservation ${S^\mu}_{;\mu}=0$ removes the dependence of one of the variables, and a particular choice of gauge eliminates another.
The resultant two radial master equations are given by Eq.~\eqref{master_eq_Psi}, which possess identical forms in terms of the variables
\begin{eqnarray}
\Psi_1(t,r) &=& a^{\ell,m}(t,r), \nonumber \\
\Psi_2(t,r) &=& \frac{r^2}{\ell(\ell+1)}\left({h^{\ell,m}}_{,0}(t,r)-{f^{\ell,m}}_{,r}(t,r)\right) ,
\label{Psi_master}
\end{eqnarray}
and the corresponding source terms are given by
\begin{eqnarray}
S_1(t,r) &=& \alpha^{\ell,m}(t,r), \nonumber \\
S_2(t,r) &=& \frac{1}{\ell(\ell+1)}\left[\left(r^2\psi^{\ell,m}\right)_{,r}(t,r)-{\eta^{\ell,m}}_{,0}(t,r)\right] .
\label{S_master}
\end{eqnarray}
The two master equations for $\Psi_{1}$ and $\Psi_{2}$ are decoupled and can be interpreted as a photon's two polarization states.
It is also noted that they correspond to given parities, $(-1)^{\ell+1}$ and $(-1)^{\ell}$, respectively.
The associated Green's functions will be denoted as $G_1$ and $G_2$.

\subsection{The Green's functions}\label{appdxB}

The arguments employed above to decompose the field in vector spherical harmonics can also be readily applied to Green's function defined by Eq.~\eqref{formal_Geq}.
In what follows, we simplify and derive the equation satisfied by Green's function Eq.~\eqref{formal_Geq}
\begin{equation}
{\mathscr{O}_x}^{\mu\rho}\left(G_{\mu\nu}(x, y)\right) = \frac{1}{\sqrt{-g}}{\delta^\rho}_\nu\delta(x-y) . \nonumber
\end{equation}
We first note that the angular decomposition for the l.h.s. of Eq.~\eqref{formal_Geq} remains unchanged for the case of Green's function.
As a result, the l.h.s. of the equation is obtained straightforwardly by replacing $\Psi^{\ell,m}_{1,2}(t, r)$ with $G^{\ell,m}_{1,2}(t, t', r, r')$ in Eq.~\eqref{master_eq_Psi}.
On the other hand, the $\delta$ function on the r.h.s. of the equation can be simplified using the completeness condition.
To be specific, for the angular part of $\delta(x-x')$, we have
\begin{eqnarray}
&& \sum_{i,\ell,m}{{\mathscr S}_{(i)}^{\ell,m}}(\theta,\varphi){{\mathscr S}_{(i)}^{\ell,m}}^\dagger(\theta',\varphi')+{{\mathscr V}_{(i)}^{\ell,m}}(\theta,\varphi){{\mathscr V}_{(i)}^{\ell,m}}^\dagger(\theta',\varphi') \nonumber \\
&&=1_{4\times 4}\ \delta(\theta-\theta')\delta(\varphi-\varphi')\frac{1}{\sin\theta} . \label{completeness_VSH}
\end{eqnarray}
It is noted that the above relation is for $4\times 4$ matrices in terms of the vector indices.
By comparing against Eqs.~\eqref{S_master} and~\eqref{formal_Geq}, one finds\footnote{In case that one might wonder the role of the implied vector index $\nu$ in ${{\mathscr V}_{(2)}^{\ell,m}}^\dagger$, 
it is simply a discrete degree of freedom associated with the point-like external source, similar to $\theta'$, a continuous degree of freedom. 
As the latter will be integrated out when the Green's function is applied to a given external source, $\nu$ will be contracted.}
\begin{eqnarray}
S^{\ell,m}_{\mathrm{G1}}&=&S^{\ell,m}_{\mathrm{G1}}(r, r', \theta', \varphi') = \frac{\delta(r-r')}{r^2}{{\mathscr V}_{(2)}^{\ell,m}}^\dagger(\theta',\varphi'), \nonumber \\
S^{\ell,m}_{\mathrm{G2}}&=&S^{\ell,m}_{\mathrm{G2}}(r, r', \theta', \varphi') = \frac{2}{\ell(\ell+1)r}\delta(r-r'){{\mathscr S}_{(1)}^{\ell,m}}^\dagger(\theta',\varphi') .\nonumber\\
\label{S_Green}
\end{eqnarray}
Subsequently, one obtains the radial sector of the equation satisfied by the Green's function
\begin{eqnarray}
\frac{\partial^2}{\partial t^2}G^{\ell,m}_{i}(t; r, r')+\left(-\frac{\partial^2}{\partial r^2}+V_\mathrm{RW}\right)G^{\ell,m}_{i}(t; r, r')\nonumber\\
=\delta(t)S^{\ell,m}_{\mathrm{Gi}}(r, r', \theta', \varphi') .
\label{master_eq_G}
\end{eqnarray}
Also, since the metric is static, the Green's function is translational invariant in time, we have replaced the temporal arguments by their difference 
\begin{eqnarray}
(t, t')\to (t-t')\to t .\nonumber
\end{eqnarray}
Due to the specific forms of Eq.~\eqref{S_Green}, we note that Eq.~\eqref{master_eq_G}, by itself, does not qualify as an equation for a univariable ``Green's function''.
Therefore, in order to solve it, one may further introduce a second Green's function $\mathcal{G}$ tailored for the radial equation.
In the frequency domain, it reads
\begin{eqnarray}
-\omega^2\mathcal{G}^{\ell,m}(\omega, r, r')+\left(-\frac{\partial^2}{\partial r^2}+V_\mathrm{RW}\right)\mathcal{G}^{\ell,m}(\omega, r, r')=\delta(r-r') .\nonumber\\
\label{master_radial_eq_G}
\end{eqnarray}
We observe it is precisely the definition of Green's function for the radial master equation of the black hole quasinormal mode.
This is because it satisfies not only the appropriate wave equation for a point source but also the outgoing wave boundary conditions.
Subsequently, the solution of the master equation Eq.~\eqref{master_eq_Psi} can be written as 
\begin{eqnarray}
\Psi^{\ell,m}_{i}(\omega; r, r' ) = \int dr' \mathcal{G}^{\ell,m}(\omega, r, r') S_i^{\ell,m}(r, r', \theta', \varphi') .\nonumber\\
\label{Psi_ell_m_i}
\end{eqnarray}
To process with Eq.~\eqref{master_eq_G}, one has
\begin{eqnarray}
G^{\ell,m}_{i}(\omega; r, r' ) = \int dr'' \mathcal{G}^{\ell,m}(\omega, r, r'') S^{\ell,m}_{\mathrm{Gi}}(r'', r', \theta', \varphi') .\nonumber\\
\label{G_ell_m_i}
\end{eqnarray}

By putting all the pieces together, we finally arrive at the formal expression for the Green's function defined in Eq.~\eqref{formal_Geq} by making use of $\mathcal{G}^{\ell,m}(\omega, r, r')$.
In particular, the Green's function for $\Psi_1$ defined in Eq.~\eqref{Psi_master} reads
\begin{eqnarray}
\widetilde G_{1}(\omega; \mathbf{r}, \mathbf{r}' ) = \sum_{\ell, m}\frac{\mathcal{G}^{\ell,m}(\omega, r, r')}{{r'}^2}{{\mathscr V}_{(2)}^{\ell,m}}(\theta,\varphi){{\mathscr V}_{(2)}^{\ell,m}}^\dagger(\theta',\varphi') .\nonumber\\
\label{G1_ell_m_i}
\end{eqnarray}
Apart from that the factor $1/{r'}^2$, the resultant expression is reminiscent of Eq.~(14) of~\cite{agr-strong-lensing-correlator-12} for the case of scalar perturbations.

Also, we clarify the notations adopted for the {\it source terms} in the above derivations.
We have introduced two different layers of Green's functions to solve the equation of motion for electromagnetic perturbations.
The term $S^\mu$ in Eqs.~\eqref{Maxwell} or~\eqref{formal_EoM} is the source term of the original Maxwell equation.
In terms of vector spherical harmonics, the equation is then simplified by separating the angular sector, and the resultant radial equations are Eqs.~\eqref{master_eq_Psi} or~\eqref{master_eq_G}.
The corresponding {\it source terms} denoted by $S^{\ell,m}$ and $S^{\ell,m}_i$ are related to $S^\mu$ and can be evaluated in terms of those of the original Maxwell equation using, for instance, the relations Eqs.~\eqref{S_master}.
When comparing Eq.~\eqref{master_eq_G} to Eq.~\eqref{master_eq_Psi}, the former is a specific case where its source terms are tailored for the Green's function defined by Eq.~\eqref{formal_Geq}.
Therefore, the subscripts of the source terms in the two cases are denoted, respectively, by $\mathrm{Gi}$ and $i$.
The radial master equations Eq.~\eqref{master_eq_G} and Eq.~\eqref{master_eq_Psi} involve two independent degrees of freedom.
However, since they possess similar form, and subsequently, identical quasinormal frequencies, they are often presented as a single equation in most literature, where the subscript $\mathrm{Gi}$ or $i$ is omitted.

\subsection{The numerical scheme to calculate the time-domain correlator}\label{appdxC}

Similar to Ref.~\cite{agr-strong-lensing-correlator-12}, one aims to evaluate the inverse Fourier transform of Eq.~\eqref{wC_Psi1} without resorting to Jordan's lemma.
The calculation consists of an integral of a strongly oscillating complex function and an infinite summation. 
To achieve this, one may utilize the freedom to choose a specific form of the emission source to their advantage.
In particular, the authors of Ref.~\cite{agr-strong-lensing-correlator-12} consider a static spherical source that possesses a specific spatial distribution.
Green's function is approximated using the asymptotic (and analytic) form of the two solutions of the corresponding homogeneous equation.
Then the inverse Fourier transform is performed for the given source.
The summation for angular momentum is carried out by truncating at a given order.

On the other hand, in Sec.~\ref{section3}, we adopt the approach introduced in~\cite{agr-qnm-lq-02} by numerically evaluating the Green's function with appropriate boundary conditions using the matrix method~\cite{agr-qnm-lq-matrix-01, agr-qnm-lq-matrix-02}.
To be specific, the main idea is first to introduce some appropriate change of variables for both the spatial coordinate and wavefunction of Eq.~\eqref{master_eq_Psi}.
As a result, the frequency-domain wavefunction is defined on the interval $\bar{x}=\frac{r-r_h}{r}\in [0,1]$ (see FIG.~\ref{Gomega}) and its boundary conditions at horizon and infinity are taken care of, similar to Leaver's continued fraction method.
The resulting second-order ordinary differential equation is then discretized and rewritten into a matrix equation, where the wavefunction and its derivatives are replaced by properly linear combinations of the (unknown) function values on the grids.
The first and last rows of the matrix are no longer physically relevant as they do not provide meaningful information on the boundary, thus replaced by $(1, 0, \cdots, 0)$ and $(0, \cdots, 0, 1)$.
Now, since the r.h.s. of Eq.~\eqref{master_eq_Psi} does not vanish, the obtained matrix is not singular, and its inverse corresponds to the Green function.
To be more specific, although the determinant of the matrix indeed vanishes when the frequency takes the (complex) values of the quasinormal frequencies $\omega=\omega_{n, \ell}$, our numerical integration only involves the real axis of $\omega$.
In other words, by inverting the matrix, one obtains the convolution of the Green's function and the external source, namely, Eq.~\eqref{wC_Psi1}.
Now, in order to facilitate the numerical integration of the frequency, we choose the form Eq.~\eqref{chi_choice}.
As explained in the text, it corresponds to a transient rather than a static source while effectively suppressing the contributions from the high-frequency part where Green's function oscillates significantly.
Also, we consider a truncation in the summation of the angular momentum.
More detailed discussions regarding the descritization process of the matrix method can be found in Refs.~\cite{agr-qnm-lq-matrix-01, agr-qnm-lq-matrix-02} and its adaptation to deal with inhomogeneous differential equation is discussed in Ref.~\cite{agr-qnm-lq-02}.

\begin{acknowledgments}
We are thankful for Dr. Sousuke Noda, who pointed us to the study of coherent autocorrelations by Chesler {\it al et.}~\cite{agr-strong-lensing-correlator-12}.
We gratefully acknowledge the financial support from
Funda\c{c}\~ao de Amparo \`a Pesquisa do Estado de S\~ao Paulo (FAPESP), 
Funda\c{c}\~ao de Amparo \`a Pesquisa do Estado do Rio de Janeiro (FAPERJ), 
Conselho Nacional de Desenvolvimento Cient\'{\i}fico e Tecnol\'ogico (CNPq), 
Coordena\c{c}\~ao de Aperfei\c{c}oamento de Pessoal de N\'ivel Superior (CAPES).
This work is also supported by National Key R\&D Program of China under contract No. 2020YFC2201400, 
Fok Ying Tung Education Foundation under contract No. 171006, 
and National Natural Science Foundation of China (NNSFC) under contract Nos. 11805166, 11775036, and 11675139.
A part of this work was developed under the project Institutos Nacionais de Ciências e Tecnologia - Física Nuclear e Aplicações (INCT/FNA) Proc. No. 464898/2014-5.
The numerical part of the research is also supported by the Center for Scientific Computing (NCC/GridUNESP) of the S\~ao Paulo State University (UNESP).
\end{acknowledgments}

\bibliographystyle{JHEP}
\bibliography{references_qian}

\end{document}